\documentclass[letterpaper]{JHEP3}
\pdfoutput=1

\usepackage{graphicx}
\usepackage{cancel}

\def\simgt{\mathrel{\lower2.5pt\vbox{\lineskip=0pt\baselineskip=0pt
           \hbox{$>$}\hbox{$\sim$}}}}
\def\simlt{\mathrel{\lower2.5pt\vbox{\lineskip=0pt\baselineskip=0pt
           \hbox{$<$}\hbox{$\sim$}}}}

\title{Mitigating Moduli Messes in Low-Scale SUSY Breaking}

\author{JiJi Fan\\
        Department of Physics, Princeton University, Princeton, NJ, 08540\\
        E-mail: \email{jijifan@princeton.edu}}
\author{Matthew Reece\\
 Princeton Center for Theoretical Science, Princeton University, Princeton, NJ, 08540\\
 E-mail: \email{mreece@princeton.edu}}
 \author{Lian-Tao Wang\\
        Enrico Fermi Institute and Kavli Institute for Cosmological Physics, University of Chicago, Chicago, IL 60637\\
        E-mail: \email{liantaow@uchicago.edu}}

\abstract{We discuss the physics of moduli (light scalar fields with Planck-suppressed couplings to matter) in the case of low-scale supersymmetry breaking such as gauge mediation. We argue that even if the mechanism of moduli stabilization is decoupled from the mechanism of SUSY breaking, moduli masses will generically be parametrically related to the gravitino mass once the cancellation of the cosmological constant is taken into account. For low-scale SUSY breaking, this implies that moduli fields are light, long-lived relics that will generically drive the universe into a matter-dominated phase, in contradiction to standard BBN. We discuss two scenarios for evading this problem. The first is to consider very tuned supergravity potentials that can make the moduli heavy enough to decay at early times and reheat above the temperature of BBN. Viable cosmology can be achieved in this scenario, which has a population of highly relativistic light gravitinos arising from decays of moduli. Next, we consider the more natural scenario with light moduli. The saxion field associated with the solution of the strong CP problem provides the most natural candidate for driving thermal inflation, which dilutes the moduli. We construct an explicit model for this scenario, when the PQ symmetry breaking scale is larger than the messenger scale. The combination of the constraints on relic abundance and gamma-ray flux from decay of the moduli favors a particular region of SUSY breaking scale, $2 \times10^3 - 10^4$ TeV. For either scenario, we find that it is generic for low-scale SUSY to be associated with late entropy production and accompanying low reheating temperatures.}

\newcommand{\beq}{\begin{equation}}
\newcommand{\eeq}{\end{equation}}
\newcommand{\beqs}{\begin{eqnarray}}
\newcommand{\eeqs}{\end{eqnarray}}

\newcommand{\calo}{{\cal{O}}}
\newcommand{\hatt}{{\hat{T}}}
\newcommand{\hatx}{{\hat{X}}}

\begin{document}

\section{Introduction}
\label{sec:intro}
\setcounter{equation}{0}
\setcounter{footnote}{0}

One of the most promising scenarios for how supersymmetry (SUSY) breaking is mediated to the Standard Model (SM) is gauge mediation (GMSB)\footnote{For a review, see the classic \cite{GiudiceRattazziReview}.}. Low-scale SUSY breaking with gauge mediation, compared to high-scale SUSY breaking with gravity mediation, enjoys two appealing features: it is automatically flavor-blind and it is a self-consistent low-energy effective field theory yielding collider-testable predictions without relying on any knowledge of quantum gravity. However, it is also well-known that the gravitino, being light and stable in low-scale SUSY breaking scenarios, is always a cosmological embarrassment~\cite{EarlyGravitino}. Specifically, for SUSY breaking scale $\sqrt{F} \in (10, 10^4)$ TeV, one has $m_{3/2} \in (100$ meV, 100 keV). If a gravitino with mass $m_{3/2} >$ keV was in thermal equilibrium at early times and froze out later, its contribution to the energy density will overclose the universe unless some late-time dilution is present. This constraint turns into an upper bound of the reheating temperature~\cite{Moroi:1993mb}. For lighter gravitinos with mass $m_{3/2} <$ keV, the overclosure constraint goes away. Yet one still needs to worry about possible limits from cosmic structure formation. The power spectrum inferred from the Ly-$\alpha$ forest data together with cosmic microwave background data of WMAP requires that the gravitino density $\Omega_{3/2} \simlt 0.12 \Omega_{DM}$, or equivalently, $m_{3/2} < 16$ eV, if the gravitino is a thermal relic decoupled at temperatures of the order of GeV or TeV~\cite{Viel:2005qj}. 

The situation is exacerbated as generally there could be other very weakly-coupled light particles present. Examples include the massless moduli that parametrize vacuum degeneracies in all known superstring theories. In this paper we will collectively refer to all light scalar fields with Planck-scale suppressed couplings to the low-scale theory as moduli, denoted by $T$, without referring to their origins. It is well known that such moduli fields can lead to serious cosmological problems, known as the moduli problem, which has received a great deal of attention in the context of gravity mediation~\cite{Coughlan:1983ci, Ellis:1986zt, deCarlos:1993jw,Banks:1993en,Randall:1994fr,Banks:1995dt,Banks:1995dp,Dine:1995uk}. When the Hubble rate drops below the mass of a modulus, the modulus starts oscillating coherently around its minimum with Planck-scale amplitude, storing an enormous amount of energy density. In gravity mediation, moduli decay around the same time as the gravitino, reheating the universe to a temperature of $T_R \sim (m_T/{\rm TeV})^{3/2}$ keV. A successful Big Bang nucleosynthesis (BBN) requires the reheating temperature to be above 5 MeV and thus the moduli masses to be above 100 TeV. 

Though moduli exist universally in both high- and low-scale SUSY breaking scenarios, they turn out to be more troublesome for the case of low-scale SUSY breaking theories, where the problem has received relatively little attention (though see Refs.~\cite{de Gouvea:1997tn,Choi:1998dw}). The reason is that generically the moduli masses are of the order of the gravitino mass if they obtain their masses after SUSY is  broken, as is often assumed. Our first goal in this paper is to clarify the connection between moduli masses and the scale of SUSY breaking. In fact, there are scenarios in which moduli stabilization is supersymmetric, e.g., by the KKLT mechanism~\cite{KKLT}. Because the dynamics stabilizing these moduli is decoupled from the dynamics breaking SUSY, superficially it would appear that the moduli problem is easily solved in such a setting. The two scales can be different, and the moduli can be made heavy enough to decay at early times, leaving conventional cosmology unscathed. However, the empirical fact that we live in a world with a very small cosmological constant (relative to the scale of SUSY breaking and other scales of particle physics) forces a relationship among {\em a priori} unrelated scales, and limits our freedom to decouple the moduli and gravitino mass scales. Without tuning, the moduli are light and long-lived. If the moduli live longer than the age of the universe, the energy density stored in their coherent oscillations exceeds the critical density of the universe, and leads to overclosure. If the moduli are unstable on cosmological timescales, they decay to two photons through the coupling $T F^2$, and the observed $\gamma$-ray backgrounds place a stringent constraint on their density~\cite{Kawasaki:1997ah,Hashiba:1997rp,Asaka:1997rv}. The various cosmological difficulties associated with gravitinos and moduli are summarized in Figure 1.

\FIGURE[!h]{
\begin{tabular}{c}
\includegraphics[width=0.62\textwidth]{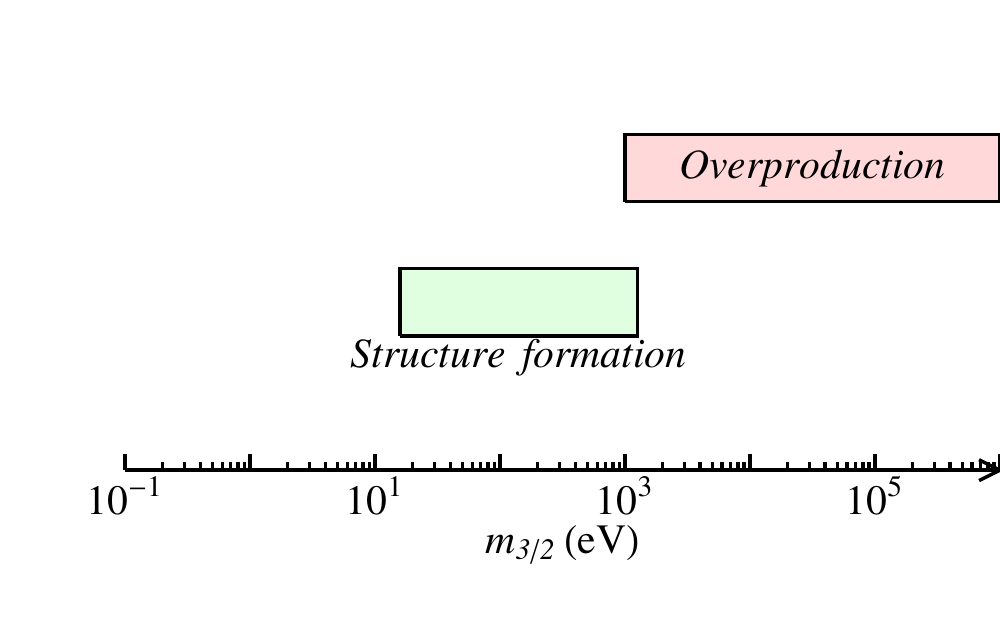} \\
\includegraphics[width=0.61\textwidth]{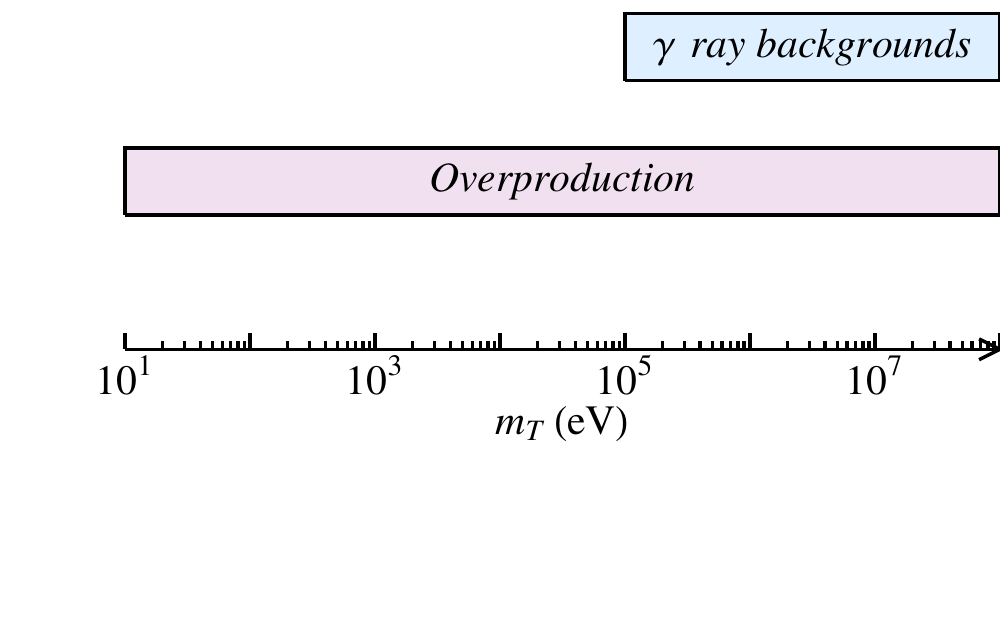}
\end{tabular}
\caption{Major cosmological constraints for light gravitino ($m_{3/2}\in (100$ meV, 100 keV)) and moduli ($m_{T} \in (10$ eV, 10 MeV)). For illustration, we assume that moduli is stabilized by KKLT-type mechanism and the moduli mass is 100 times as large as the gravitino mass. Yet it should be borne in mind that the exact relation between the two masses is model dependent. }
\label{fig:gravitino}}

In this paper, we will investigate two possible methods to alleviate the cosmological problems of low-scale SUSY breaking. First, it is possible, via tuning the high-scale moduli superpotential, to give the moduli large supersymmetric masses, making them parametrically much heavier than the gravitino, e.g., $m_T$ could be above 100 TeV. They would be unstable, decaying to gravitinos, SM particles, and superpartners at a temperature above 5 MeV, as required by BBN. Primordial gravitinos will be diluted while the moduli oscillate, and if the reheating temperature is below a TeV, gravitino overclosure will not be a problem~\cite{Moroi:1993mb}. However, moduli decays $T \to 2\psi_{3/2}$ can be a new source of additional {\em relativistic} gravitinos, which may spoil the success of standard BBN. We will discuss this decay in some detail.

In the case that not all moduli are tuned to be heavy by some high-scale dynamics, a new mechanism needs to be in place to produce additional entropy to dilute the moduli significantly after they start oscillating (while at the same time diluting the gravitino). A period of late-time mini-inflation is frequently invoked for this purpose~\cite{Randall:1994fr}. Thermal inflation~\cite{LythStewart} provides an appealing example, with a light field trapped at a fixed value by thermal corrections driving the inflation. It requires a scalar with a very flat potential, which is already present for a field in gauge mediation with a strong CP solution: the saxion! We argue that it is natural to connect the solution to moduli problems with the solution to the strong CP problem. Through mixing with the SM Higgses, the saxion would decay to SM particles which thermalize at a low reheating temperature without spoiling BBN. In contrast to a model proposed recently \cite{Choi:2011rs}, we consider the case where the PQ symmetry breaking scale, typically constrained to be $> 10^9$ GeV, is much higher than the messenger scale of the low energy SUSY breaking. After thermal inflation, the moduli density scales as $T_{RH}(T_{\rm roll}/T_{\rm T.I.})^3$ where the saxion reheating temperature has to take the lowest value required by BBN $T_{RH}\sim$ 10 MeV to allow for enough dilution. The temperature where thermal inflation ends is set by soft masses and hence fixed around the weak scale $T_{\rm roll} \sim \calo(100$ GeV - 1 TeV). Thus the only free parameter left is the thermal inflation starting temperature, which gives the most dilution at larger values of the thermal inflaton potential height, $T_{\rm T.I.} \propto \sqrt{F}$. On the other hand, smaller $\sqrt{F}$ implies lighter moduli and longer moduli lifetimes $\tau \propto F^{-3}$, which are favored by the $\gamma$-ray data~\cite{Kawasaki:1997ah,Hashiba:1997rp,Asaka:1997rv}. Thus there is a very strong and generic tension between achieving enough dilution of the moduli through thermal inflation, which favors larger values of $\sqrt{F}$, and avoiding the constraints on decaying moduli, which favors smaller values of $\sqrt{F}$ (and disfavors the model of Ref.~\cite{Choi:2011rs}). Although there is enough model-dependence in the moduli sector that it is difficult to make completely definitive statements, we will argue that successfully diluting moduli, achieving a high enough reheating temperature for BBN, and avoiding constraints from $\gamma$-rays selects a very particular parameter range, with $\sqrt{F} \approx 2 \times10^3~{\rm to}~10^4$ TeV, $f_a \approx 10^{11}~{\rm to}~10^{12}$ GeV, and moduli masses in the vicinity of the gravitino mass rather than heavier as in KKLT.

The paper is organized as follows. In Sec.~\ref{sec:problem}, we will first review the KKLT mechanism and show that even if moduli are stabilized supersymmetrically, their masses remain of order $m_{3/2}$. Then we demonstrate that tuning (separate from the tuning that cancels the cosmological constant!)~can stabilize the moduli with large supersymmetric masses. After that, the cosmological properties of the heavy moduli are studied. In Sec.~\ref{sec:saxion}, we turn to the saxion solution to the light moduli problem. We present the saxion properties and the parameter space where the thermal inflation followed by the saxion's oscillation around its minimum and decays would produce enough entropy to dilute the moduli density. We discuss other issues and conclude in Sec.~\ref{sec:discuss}. In the appendix, we give a brief review of the origin and properties of moduli fields.

\section{The Moduli Problem and the Tuning Solution}
\label{sec:problem}
\setcounter{equation}{0}
\setcounter{footnote}{0}

We begin by considering some simple models in supergravity that illustrate properties of moduli stabilization coupled to a SUSY breaking sector. Because moduli masses arising {\em from} SUSY breaking are necessarily of order $F/M_P$ (since moduli couplings are Planck-suppressed), we are most interested in cases in which moduli stabilization is decoupled from SUSY breaking at leading order, which at least offer the possibility of lifting the moduli masses to much higher scales. The two models we discuss appeared in Refs.~\cite{OKKLT,AbeMore}, but there the emphasis was on gravity mediation and the relevance to the moduli problem was less apparent. Here we discuss the results with an emphasis on scenarios with very light gravitinos.

We set $M_P=1$ in most intermediate steps but put it back to clarify results where numerical orders of magnitude are important. We use the reduced Planck mass $M_P = 2.4\times 10^{18}~{\rm GeV}$, for which the gravitino mass is $m_{3/2} = \frac{F}{\sqrt{3} M_P}$ and the critical density is $\rho_c = 3 H^2 M_P^2$.

\subsection{A Minimal Model: O'KKLT}

We begin with a simple example to show how the scales in the moduli sector and the low-scale SUSY breaking sector are tied up. We couple the KKLT model~\cite{KKLT} to the simplest possible SUSY breaking sector, a Polonyi model with an added K\"ahler potential term to lift the flat direction. This model is well-studied under the name O'KKLT~\cite{OKKLT}, and closely related work appears in Refs.~\cite{LutySundrum,OKKLTlike}. The model is 
\beqs
K_{O'KKLT} & = & - 3 \log(T+T^\dagger) + X^\dagger X - \frac{\left(X^\dagger X\right)^2}{2M^2}, \nonumber \\
W_{O'KKLT} & = & W_0 + \Lambda^3 e^{-bT} + f X.
\eeqs
We will assume $\Lambda \sim M_P$ (or another very high scale).  The constant $W_0$ originates from fluxes in string compactification which can be tuned to make $W_0 \ll M_P^3$. The dimensionless coefficient $b$ is $2 \pi$ in the case of D3-brane instantons and $2\pi/N$ if it arises from gaugino condensation on D7 branes. For a discussion of the physics behind the logarithmic form of the K\"ahler potential, see Appendix~\ref{app:modprop}.~The $X$ sector is just the Polonyi model of SUSY breaking with a K\"ahler correction that stabilizes $\left<X\right>$ at the origin. The cutoff scale $M$ in the $X$ sector is related to the underlying strong dynamics that breaks SUSY, and could be much smaller than $M_P$.

As the two sectors are only connected by gravity, we can find the minimum of the potential by first inspecting the two sectors separately. The KKLT sector possesses a supersymmetric AdS minimum at large modulus value\footnote{Here ${\rm W}_{-1}$ denotes the branch of the Lambert W-function (implicitly defined by $z = {\rm W}e^{\rm W}$) which is well-defined for small negative arguments $\geq -1/e$ and takes values $\leq -1$ \cite{LambertW}.}
\beqs
&& D_T W \equiv W_T+K_T W= 0 \nonumber \\
&\Rightarrow& bt_* = -{\rm W}_{-1}\left(-e^{3/2}\frac{3W_0}{2\Lambda^3}\right)+\frac{3}{2} \sim \log\frac{\Lambda^3}{\left|W_0\right|},
\eeqs
with potential depth
\beq 
V_{AdS} = -\frac{b^2 \Lambda^6 e^{-2 bt_*}}{6 M_P^2 t_*} \sim -\frac{W_0^2}{M_P^2}.
\eeq
After SUSY breaking, the $X$ sector will contribute to the potential $V_X \sim |f|^2$. The vacuum energies of opposite signs from the two sectors have to be balanced against each other, $f \sim \left|W_0\right|/M_P$, to obtain an almost vanishing c.c. The modulus mass is determined by $W_0$, and thus is of the same scale as the AdS curvature and the gravitino mass. More quantitatively, perturbing around the minimum of two disconnected sectors, one finds that, up to corrections scaling as $1/(bt_*)$, 
\beqs
V|_{min} \approx 0 &\Rightarrow& f=\frac{\sqrt{3} |W|}{M_P} \nonumber \\
m_{3/2} &=& e^{G/2}= \frac{f}{2\sqrt{6}t_*^{3/2}M_P}, \\
m_{T}&=& e^{G/2}G^{TT^\dagger}G_{TT}\sim m_{3/2} \log\frac{\Lambda^3}{\left|W_0\right|},
\eeqs
where the total K\"ahler potential $G\equiv K+\log |W|^2$. Indeed the modulus is heavier than the gravitino, but only by a logarithmic factor. A recent study of supersymmetric sigma models in AdS has shown that generically, a theory with many KKLT-like moduli with potential dominantly arising from one exponential in the superpotential will have all moduli masses proportional to the AdS curvature scale, times a logarithmic factor~\cite{Adams:2011vw}.

\FIGURE[!h]{
\includegraphics[width=0.45\textwidth]{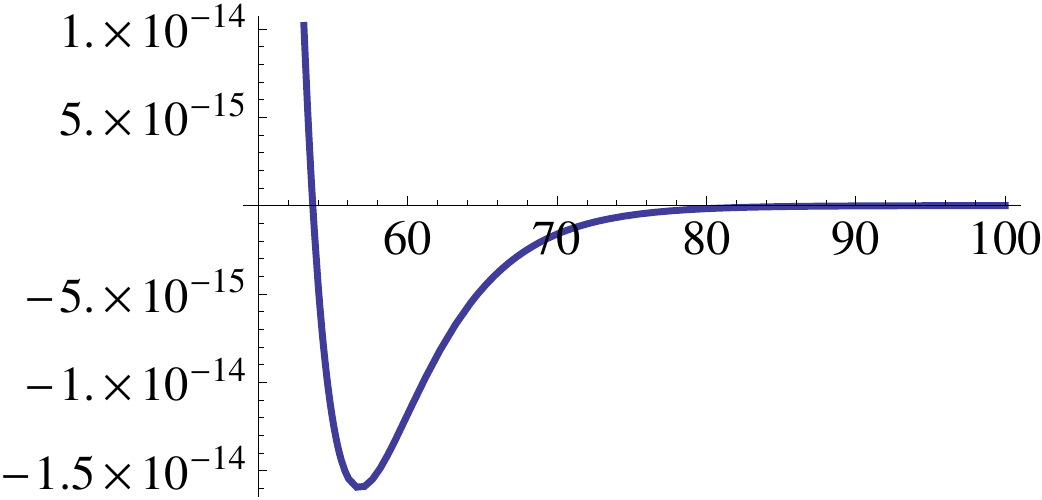}
\caption{An example KKLT potential $V(T)$, for parameter choices leading to a minimum at $t_* = 56.8$ with depth $V(t_*) = -1.4 \times 10^{-14}$.}
\label{fig:KKLT}}

\subsection{A Tunable Superpotential}
\label{sec:tunable}
Now we will demonstrate that the modulus can be made parametrically much heavier than the gravitino through tuning the modulus superpotential. The model appears already in Refs.~\cite{OKKLT, AbeMore}. Keeping the same K\"ahler potential $K_{O'KKLT}$, consider a racetrack type superpotential with two exponential terms in the moduli sector (plus a constant): 
\beq
W_{tunable} = W_0 + \Lambda^3 e^{-b T} - \Lambda^{^\prime3} e^{-b^\prime T}.
\label{eq:Wtunable}
\eeq
Before discussing the detailed physics, let us outline the general strategy: we want to have a vacuum where the moduli are much heavier than the gravitino. Moduli masses come from derivatives of the superpotential $W$ at the minimum, while the gravitino mass comes from $\left<W\right>$. Thus, we want to tune so that at the minimum $\left<W\right>$ is small compared to the terms that make it up. One way to do this is to first tune to produce a supersymmetric Minkowski vacuum, then perturb it to produce a supersymmetric AdS vacuum for the modulus, then couple to $X$ to uplift.

All three terms in the superpotential are taken to be comparable, $W_0, \Lambda^3, \Lambda^{\prime 3} \sim \calo(M_P^3)$. If any of them is much smaller than the others, it would not influence the AdS minimum, the modulus mass would still be of order the AdS curvature, and the physics would resemble that discussed in the previous subsection. The addition of the third term of the same order allows us to tune the parameters $W_0, \Lambda, \Lambda', b,$ and $b'$ to produce a supersymmetric \textbf{Minkowski} minimum by setting $D_T W =0$ and $W=0$, 
\beqs
t_*^{(0)} &=& \frac{1}{b-b^\prime} \log \frac{b \Lambda^3}{b^\prime\Lambda^{\prime3}},\label{eq: Minmin} \\
W_0^{(0)} &=& \Lambda^{\prime3}\left(\frac{b\Lambda^3}{b^\prime\Lambda^{\prime3}}\right)^{\frac{b^\prime}{b^\prime-b}}\left(1-\frac{b^\prime}{b}\right).
\label{eq: MinminW0}
\eeqs
For a controlled solution, the modulus should still be stabilized at large volume $t_*^{(0)} \gg 1$. This could be achieved, e.g.,  by making $\Lambda$ relatively larger than $\Lambda^\prime$. Notice that this is quite different from the KKLT scenario, where the large volume stabilization is obtained through tuning $W_0$ to be small. Another crucial difference is that though the gravitino is still massless, the modulus has  already obtained a large supersymmetric mass
\beq
m_T=\frac{\sqrt{2t_*^{(0)}}\Lambda^3}{3M_P^2}e^{-bt_*^{(0)}}b(b-b^\prime). 
\label{eq: tunedmodulimass}
\eeq

Next we couple the racetrack sector to the SUSY breaking sector $fX$ with $f \ll m_T M_P$. However, it is easier to first ignore $X$ and build a supersymmetric AdS minimum for the pure modulus sector by shifting the parameter $W_0$ so that at the new minimum $3\left|W\right|^2 = f^2$, while the condition $D_T W = 0$ is preserved. This leads to the following constraints on the parameters:
\beqs
W_0 + \Lambda^3 e^{-b t_*} - \Lambda^{^\prime3} e^{-b^\prime t_*}  &=&\frac{fM_P}{\sqrt{3}},\nonumber \\
-b\Lambda^3 e^{-b t_*} +b^\prime\Lambda^{^\prime3} e^{-b^\prime t_*}&=&\frac{\sqrt{3}fM_P}{2t_*}.
\eeqs
Thus, the modulus sits in a supersymmetric \textbf{AdS} minimum with the AdS curvature determined by the perturbation. Compared to Eq.~\ref{eq: Minmin}, the position of the minimum shifts, between the SUSY Minkowski minimum and the SUSY AdS minimum, but only by a suppressed amount $\delta t_* \sim f/(M_P m_T)$ and the solution $W_0^{(0)}$ shifts to
\beq
W_0=\Lambda^{\prime3}\left(\frac{b\Lambda^3}{b^\prime\Lambda^{\prime3}}\right)^{\frac{b^\prime}{b^\prime-b}}\left(1-\frac{b^\prime}{b}\right)+\frac{fM_P}{\sqrt{3}}.
\label{eq:tuning}
\eeq
At the AdS minimum, the gravitino mass, of order AdS curvature, is also set by the perturbation $f$,
\beqs
m_{3/2}=\frac{f}{\sqrt{3}M_P}\frac{1}{(2t_*)^{3/2}} \, , 
\eeqs
which is parametrically much smaller than the modulus mass in Eq.~\ref{eq: tunedmodulimass}. Having constructed this AdS minimum, we can then add the $X$ field to uplift it to a Minkowski minimum, which, similarly to the uplifting of KKLT in the previous subsection, has only a small effect on the value of $t_*$ at the minimum, as the two sectors are nearly decoupled.

Before ending this section, we would like to summarize the differences between the tuned case and the KKLT case in the previous section. In the tuned case, $|W_{tunable}|$ is much smaller than each of its individual terms. The modulus mass is determined by the supersymmetric parameters, dissociated from the SUSY breaking order parameter $f$ that sets the gravitino mass. From Eq.~\ref{eq:tuning}, one could see that the tiny c.c. is achieved by a fine tuning to the order $\calo(fM_P/\Lambda^3)$ among the large Planck-scale parameters in the moduli sector. The modulus potential is shallow but steep as shown in the right panel of Figure~\ref{fig:Racetracks}. We stress that the tuning that makes the modulus heavy is {\em different} from the tuning that cancels the c.c.; even fixing the depth of the supersymmetric AdS minimum that we uplift, the typical potential will look like the left-hand plot in Fig.~\ref{fig:Racetracks}, and the modulus mass will be of order $m_{3/2}$.

\FIGURE[!h]{
\includegraphics[width=0.4\textwidth]{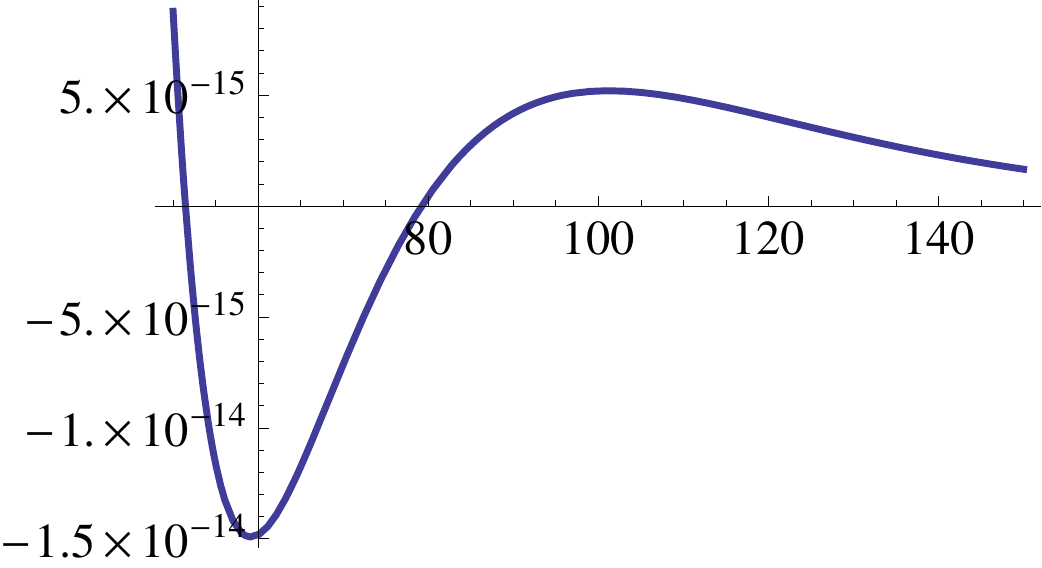}~~\includegraphics[width=0.4\textwidth]{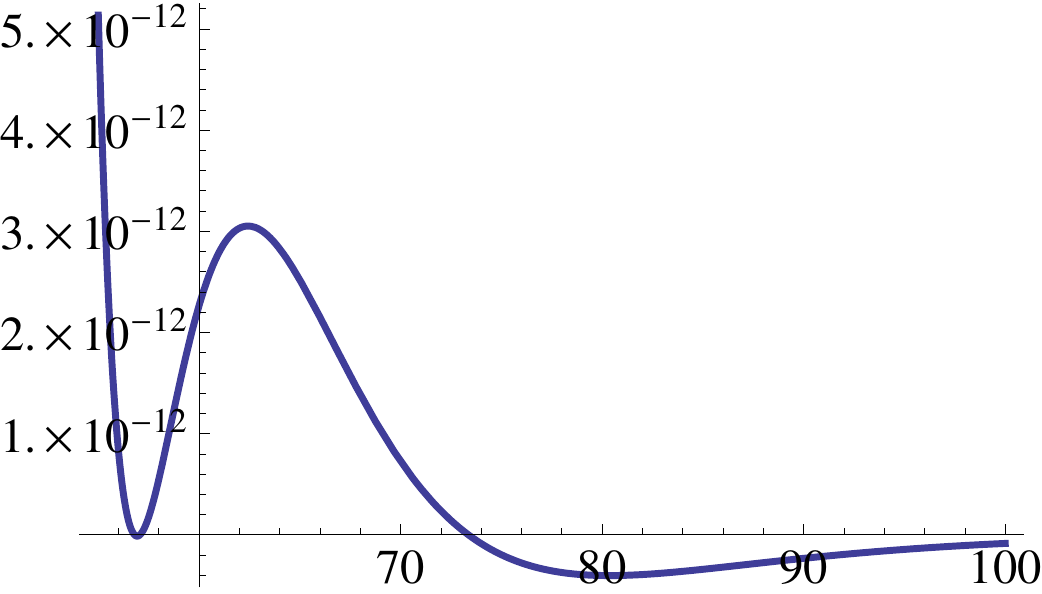}
\caption{Two example potentials $V(T)$, untuned at left and tuned at right, arising from the superpotential \ref{eq:Wtunable}. As in Fig. \ref{fig:KKLT}, there is a minimum at $t_* = 56.8$ with depth $V(t_*) = -1.4 \times 10^{-14}$ in each case. In the plot at right, the modulus at this minimum is tuned to be very heavy.}
\label{fig:Racetracks}}

Our argument that moduli masses will be of order the gravitino mass unless there is a tuning such that the individual terms in $W$, at the minimum, nearly cancel is very similar to the one already given in Ref. \cite{deCarlos:1993jw}. However, they explicitly state an assumption that moduli fields are massless in the absence of SUSY breaking. We emphasize the key point that we {\em need not} assume that the moduli masses arise from SUSY breaking effects to make this argument.

\subsection{Moduli Decays}
If we tune to have a light gravitino and heavy moduli, one potential problem is an overabundance of gravitinos arising from moduli decays. The heavy moduli could either cascade through SM superpartners ending in gravitinos and SM particles, e.g., $T \to 2\tilde{g} \to \cdots \to  2\psi_{3/2}+ \cdots$ or decay directly to a pair of gravitinos $T \to 2\psi_{3/2}$. The decay widths scale as $\Gamma_T = \alpha m_T^3/(4\pi M_P^2)$ with a model-dependent constant $\alpha$. Below we will calculate the partial widths and the branching fractions of different decay channels.

The simplest way to couple a single modulus to the MSSM is through the gauge kinetic function, $\int d^2 \theta \,T W^\alpha W_\alpha$, which leads to, in components, 
\beq
{\cal{L}} = -\frac{1}{4\sqrt{G_{TT^\dagger}}}\left( \frac{\hat{T}_R}{t_*}G^{a}_{\mu\nu}G^{a\mu\nu}+\frac{\hat{T}_I}{t_*}G^{a}_{\mu\nu}\tilde{G}^{a\mu\nu} - \frac{m_T}{t_*} (\hat{T}\lambda \lambda + h.c.) \right),
\eeq
where the subscripts $R(I)$ denote the real (imaginary) component of $T$, and $\hat{T} \equiv \sqrt{G_{TT^\dagger}} (T- t_*) $ is the canonically normalized modulus field. For $K=-3\log(T+T^\dagger)$, a straightforward calculation yields the partial widths 
\beq
\Gamma(T_{R,I} \to g g) = \Gamma(T_{R, I} \to \lambda \lambda) = \frac{N}{96 \pi }\frac{m_T^3}{M_P^2},
\label{eq:cascade}
\eeq
where $N$ counts the degrees of freedom of the final states. For instance, $N=8$ for gluons and gluinos. The decay widths to the gauge bosons and gauginos are equal as SUSY breaking effects are suppressed by $m_{3/2}/m_T$ or $m_{soft}/m_T$. Notice that the result is independent of the gauge coupling or the modulus VEV, which is generically not true when the MSSM is coupled to multiple moduli. Other couplings such as $\int d^4 \theta Q^\dagger e^{-g(T)V} Q$ would induce three-body decays such as $T \to \tilde{q} \bar{\tilde{q}} g$ but these would be suppressed by the gauge coupling as well as a three-body phase space factor. Thus we will neglect them for the rest of the discussion. 

The direct decay of moduli to gravitinos, $T \to 2\psi_{3/2}$, is one potential source of cosmological difficulties~\cite{ModuliToGravitinos}. The coupling here depends on the SUSY-breaking $F$-term of the modulus field. The generic expectation is that for heavy moduli $G_\hatt \sim m_{3/2}/m_T$. Let us briefly review the argument. Suppose, around the supersymmetric vacuum, we had a large mass, $\frac{1}{2} m_T^2 (T - t_*)^2$. After coupling to the SUSY-breaking field, we have a new term in the potential, $e^K K^{X^\dagger X} \left|f\right|^2$. Because $e^K$ is $T$-dependent, the VEV of $T$ will shift, between the supersymmetric AdS vacuum and SUSY-breaking Minkowski vacuum, by $\delta t_* \sim \frac{|f|^2}{m_T^2}$. This leads to a new $F$-term for $T$, $F_\hatt \sim \delta t_*~\partial_\hatt(D_\hatt W) \sim \frac{|f|^2}{m_T}$. Thus, we conclude that $G_\hatt \sim m_{3/2}/m_T$, in which case the decay width is order $m_T^3/M_P^2$~\cite{ModuliToGravitinos}.

However, an important subtlety arises in calculating this decay~\cite{Dine} (see also the generalizations in Ref.~\cite{Endo:2006tf}). The physical Goldstino is a linear combination of $\psi_X$ and $\psi_T$ while the orthogonal combination $\Psi$ is massive. Similarly, its massive scalar partner $\Phi$ is also a linear combination of $X$ and $T$. It was first pointed out in~\cite{Dine} that in certain classes of models, such as KKLT, the mixing is supersymmetric at $\calo(m_{3/2}/m_T)$. Consequently the decay amplitude of the modulus, or more exactly of the massive scalar $\Phi$, starts at $\calo((m_{3/2}/m_T)^2)$. In the KKLT model, the two small numbers $m_{3/2}/m_T$ and $1/t_*$ are of the same order, but this not true in more general models like that of Sec.~\ref{sec:tunable}, for which $m_{3/2}/m_T \ll 1/t_*$. In order to disentangle which small factor appears in the general form of the argument of Ref.~\cite{Dine}, and for completeness of the discussion, we will go into more details of the argument below.

The coupling of a heavy scalar $\Phi$ to the gravitino is $e^{G/2}G_\Phi \Phi\bar{\psi}_\mu\sigma^{\mu\nu}\psi_\nu/2$. Thus, to calculate the width of $\Phi$, one needs to work out $G_\Phi$. In terms of the $G$ function, the potential is $V=e^G(G^{ii^\dagger}G_iG_{i^\dagger}-3)$, where $i$ runs over all the fields. To have vanishing c.c., $G^{ii^\dagger}G_iG_{i^\dagger}-3=0$. The calculations below will assume a factorizable K\"ahler potential $K=-3\log(T+T^\dagger)+k(|X|^2)$ where $k(|X|^2)$ is a function of $|X|^2$. From the stationary condition $\partial V/\partial T^\dagger =0$, one obtains 
\beqs
G_T&\equiv&\partial_T G=-\frac{G^{XX^\dagger}G_{X^\dagger T^\dagger}}{G^{TT^\dagger}G_{T^\dagger T^\dagger}}G_X\left(1+\calo(m_{3/2}/m_T)\right) \nonumber \\
&=& -\frac{m_{3/2}}{m_T}G^{XX^\dagger}G_{X^\dagger T^\dagger}G_X \left(1+\calo(m_{3/2}/m_T)\right),
\eeqs
where we employ $m_T= e^{G/2}G^{TT^\dagger}G_{TT}=m_{3/2}G^{TT^\dagger}G_{TT}$ in the second line. Next we will diagonalize the scalar mass matrix. The off-diagonal term would induce a mixing between $T$ and $X$, 
\beq
V_{\hatt \hatx} = \frac{G_{X^\dagger T^\dagger}}{G_{T^\dagger T^\dagger}} \sqrt{\frac{G^{XX^\dagger}}{G^{TT^\dagger}}}m_T^2,
\eeq
where $\hatx$ is canonically normalized.
Thus the heavy mass eigenstate is, assuming $m_T > m_X$, 
\beq
\Phi = \epsilon \hatx+\hatt,
\eeq
where
\beqs
\epsilon &=& \frac{m_T^2}{m_T^2-m_X^2} \frac{G_{X^\dagger T^\dagger}}{G_{T^\dagger T^\dagger}} \sqrt{\frac{G^{XX^\dagger}}{G^{TT^\dagger}}} \nonumber \\
&=&\frac{m_T^2}{m_T^2-m_X^2}\frac{m_{3/2}}{m_T}\sqrt{G^{XX^\dagger}G^{TT^\dagger}}G_{X^\dagger T^\dagger}.
\eeqs
The factor relevant for the decay, $G_\Phi$, is (at leading order in $m_{3/2}/m_T$) 
\beqs
G_\Phi &=& \epsilon G_\hatx + G_\hatt = \epsilon G_\hatx+\sqrt{G^{TT^\dagger}}G_T \nonumber \\
&=&\left (\frac{m_T^2}{m_T^2-m_X^2} -1\right)\frac{m_{3/2}}{m_T}\sqrt{G^{TT^\dagger}}G^{XX^\dagger}G_{X^\dagger T^\dagger}G_X \\
&=&3\sqrt{3} \left (\frac{m_X^2}{m_T^2-m_X^2} \right)\frac{m_{3/2}}{m_T}G^{XX^\dagger}. \label{eq:suppresseddecay}
\eeqs
In the limit $m_X \ll m_T$, there is a suppression of the direct decays. The argument of Ref.~\cite{Dine} has a ``chirality-suppressed" coupling $G_\Phi \sim \left(m_{3/2}/m_T\right)^2$ in KKLT coupled to a flat pseudomodulus $X$, matching the $m_X \to 0$ limit of Eqn.~\ref{eq:suppresseddecay}. However, in the low-scale SUSY breaking limit we are interested in, where the mass of $X$ is present in effective field theory in the $M_P \to \infty$ limit, we will generally have $m_X^2/m_T^2 \gg m_{3/2}/m_T$ and Eqn.~\ref{eq:suppresseddecay} is the dominant term. For generic low-scale SUSY breaking, with completely general K\"ahler potential, any suppression beyond $G_\Phi \sim m_{3/2}/m_T$ is absent, leading to a decay width
\beq
\Gamma(T \to 2\psi_{3/2}) = \frac{c}{288\pi} \frac{m_T^3}{M_P^2},
\label{eq:directdecay}
\eeq
where $c$ is a model-dependent order-one factor. For the nongeneric form $K=-3\log(T+T^\dagger)+k(|X|^2)$, $c=27m_X^4/(k_{XX^\dagger}(m_T^2-m_X^2)^2)$. 

\subsection{Cosmological constraints}
Heavy moduli have a relatively high reheating temperature 
\beqs
T_R &=& (\pi^2 g_{*s}/90)^{-1/4}\sqrt{M_P \Gamma_T} \nonumber \\
&\approx& 5.5 \times 10^{-3} {\rm{MeV}}\left( \frac{m_T}{{\rm{ 1\,TeV}}}\right)^{3/2},
\eeqs
where we approximated the modulus decay width by $\Gamma_T = m_T^3/(4\pi M_P^2)$. For $m_T >$ 100 TeV, $T_R \gtrsim 5$ MeV, compatible with the requirement for a successful BBN.\footnote{A related discussion of the need for tuning to make moduli heavy appeared recently in Ref.~\cite{deAlwis:2010ud}, which argued that moduli should be made heavier than the {\em messenger} scale so that they can be integrated out. We will not impose such strict demands; because moduli couple very weakly, they don't affect the dynamics of the SUSY-breaking sector in detail, so we only consider phenomenological and cosmological constraints on their masses. These considerations only require that they be heavier than 100 TeV.} In the range $\sqrt{F} \in (10^3 - 10^4)$ TeV, $m_{3/2} \in (1 - 100)$ keV,  the decay of the heavy moduli with mass $m_T \in (100 - 10^6)$ TeV could reduce the gravitino density below the critical density. However, one still needs to worry about the gravitinos from the direct decay of the moduli $T \to 2\psi_{3/2}$. The energy they carry could increase the Hubble expansion rate, resulting in an overproduction of He$^4$. More quantitatively, the energy density of the gravitinos has to be smaller than $20\%$ of that of the SM particles, or equivalently, $Br(X \to 2 \psi_{3/2})< 0.2$. Combining Eq.~\ref{eq:cascade}, ~\ref{eq:directdecay}, for the tuned model we consider, this constraint turns into $c<18$, which could be easily satisfied for $m_X < m_T$ in the case of a nongeneric K\"ahler potential. For a completely generic K\"ahler potential, however, this moduli-induced gravitino problem is a dangerous constraint on the tuned scenario~\cite{ModuliToGravitinos}. 

 One might also worry about whether similar problems would be caused by gravitinos at the end of the cascade decay chains of moduli. In this case, however, the superpartners produced from the decaying moduli would first enter the thermal bath, leaving no chance for them to decay to gravitinos as the thermalization time scale $\calo$(GeV$^{-1})$ is much shorter than the lifetime of the decay to gravitino $\sim F^2/m_{soft}^5 \gg$ $\calo$(GeV$^{-1})$. Yet, eventually the next-to-the lightest superparticles (NLSPs) will freeze out, and their late decays to the gravitino after the BBN epoch would modify the abundances of the light elements and thus are constrained~\cite{Moroi:1993mb}. For instance, for $\tilde{\tau}$ as NLSP, its lifetime has to be $\Gamma^{-1} < 6 \times 10^3$ sec~\cite{PospelovBBN}. This is easily satisfied for low-scale gauge mediation.

We close this section by noting that, if all moduli are tuned to be heavy, any QCD axion in the theory should not be a fundamental (e.g., string theory) modulus, but rather a conventional field coupling with larger than gravitational strength. In particular, because the QCD axion should get its potential dominantly from QCD instantons, in the limit of unbroken SUSY the QCD saxion will also be a flat direction, and the saxion mass scales as $F/M_P$ if it is a fundamental modulus. Thus, the QCD saxion can never be tuned to be heavy without ruining the solution of the strong CP problem. (General comments on axions as unstabilized fundamental moduli may be found in Ref.~\cite{AxionsInStringTheory}.) A field with couplings that are not Planck-suppressed, however, can play the role of the QCD axion, as the stronger couplings can allow the saxion to get a larger SUSY-breaking mass. Thorough recent discussions of axion and saxion couplings and cosmology in the limit when moduli are assumed to be decoupled (as in this tuned scenario) appear in Ref.~\cite{AxionsInFieldTheory}. However, given that any theory with low-scale SUSY breaking and a QCD axion should contain a saxion field that is not a Planck-coupled modulus, we find it more compelling to avoid tuning, keep the moduli light, and use the saxion dynamics to dilute their abundance. We now turn to this scenario in the next section.

\section{The Saxion Scenario}
\label{sec:saxion}
\setcounter{equation}{0}
\setcounter{footnote}{0}

\subsection{The Model and the PQ Scale / SUSY Scale Relation}

In this section we will discuss what we consider to be a natural solution to the moduli problem in the gauge mediation context: namely, diluting the moduli through cosmological dynamics driven by the saxion field associated with the solution of the strong CP problem. This will be a concrete example of the general paradigm of thermal inflation, as developed by Lyth and Stewart~\cite{LythStewart}, and originally applied to low-scale gauge mediation in Refs.~\cite{de Gouvea:1997tn,Choi:1998dw}. By tying it to other physics, we aim to make the general picture more concrete and palatable. Closely related recent work by Choi et al. appeared while this paper was in preparation~\cite{Choi:2011rs}, although we will mostly focus on a different part of the parameter space for the model.\footnote{A preliminary version of our work was presented earlier \cite{ReeceDavisTalk}.} Another similar model, with an emphasis on the $\mu/B\mu$ problem, appeared recently in Ref.~\cite{Jeong:2011xu}. Related work in the gravity mediation context can be found in Ref.~\cite{MoreThermalInflation}.

We consider a model with two singlets $S$ and $Y$ and vectorlike quarks $Q_1, Q_2$ with superpotential
\beq
W = \lambda_Q S Q_1 Q_2 + \lambda_H \frac{S^2}{M_P} H_u H_d + \lambda_Y \left(\frac{S}{M_P}\right)^n S^2 Y + W_{GMSB},
\label{eq:saxionsuperpotential}
\eeq
where we will focus on cases with $n=1,2$. $W_{GMSB}$ includes the MSSM superpotential (except for the $\mu$-term, which is supplied by $\lambda_H$ above), interactions with messengers, the SUSY-breaking sector, and so on. This model is a version of the Kim--Nilles approach to the $\mu$ problem, linking the PQ and TeV scales \cite{KimNilles}. In the supersymmetric limit, the $|F_Y|^2$ term in the potential causes the dynamics to favor $\left<S\right> = 0$. On the other hand, after SUSY breaking, gauge mediation will affect the $Q_i$ fields, which in turn affect $S$. 

One desideratum of an axion model is to have $10^9~{\rm GeV} \simlt f_a \simlt 10^{12}~{\rm GeV}$, where the lower bound comes from stellar cooling and the upper bound from requiring that the axion relic abundance does not overclose the universe. (See Ref. \cite{KimCarosiReview} for a recent review of axion physics.) In our model, we should make the identification:
\beq
f_a \equiv \sqrt{2} \left<S\right>,
\eeq
because after integrating out the heavy quarks $Q_1, Q_2$, we find an induced coupling:
\beq
\int d^2\theta \frac{\alpha_s}{8\pi}\log S~{\rm tr}W_\alpha W^\alpha + h.c.,
\eeq
and using $S=(\left<S\right>+\frac{s}{\sqrt{2}})e^{i\frac{a}{\sqrt{2}(\left<S\right>}}$, we find that this includes:
\beq
{\cal L}_{\rm axion} \supset \frac{\alpha_s}{8\pi f_a} a F^a_{\mu\nu}\tilde{F}^{a\mu\nu},
\eeq
with $f_a =\sqrt{2} \left<S\right>$, which matches the standard normalization of the axion--gluon--gluon coupling. 

As with any field theory model that aims to solve strong CP, the expectation that gravity breaks continuous global symmetries will require that Planck-suppressed PQ-violating operators added to the superpotential~\ref{eq:saxionsuperpotential} or the K\"ahler potential are forbidden by a discrete symmetry~\cite{Georgi:1981pu,MorePQwithGravity}. More specifically, take an additional PQ breaking operator in the superpotential $S^{m+3}/M_P^m$ as an example. Its contribution to the axion potential is $\delta V \sim f_a^{m+2}/M_P^{m} |F_S| \sim f_a^{m+3} v^2/M_P^{m+1}$ with $v=246$ GeV. For $f_a = 10^9~(10^{12})$ GeV, this would overcome the QCD contribution $\bar{\theta}m_\pi^2f_\pi^2$ unless $m>3~(6)$, corresponding to $N> 6~(9)$ for an underlying discrete ${\mathbb Z}_N$ symmetry. 

For concreteness, we will discuss the case of minimal gauge mediation; i.e., we will assume a field $X$, $\left<X\right> = M_{mess} + \theta^2 F$, with superpotential coupling $X\overline{\Phi}\Phi$ to messengers in the ${\bf 5}$ and ${\bf{\bar 5}}$ of SU(5). Extending the discussion to the more general case \cite{GGM} is straightforward. At the messenger scale, the soft mass $m_S^2 = 0$ (up to threshold corrections), whereas $Q_{1,2}$ get positive squared masses from GMSB. In the minimal case we have (neglecting effects of RG running, and assuming QCD couplings dominate):
\beq
m_{Q_i}^2 = \frac{\alpha_s^2}{6\pi^2} \frac{F^2}{M_{mess}^2}.
\label{eq:MGMsquarkmass}
\eeq
The usual one-loop RGE for the soft masses, 
\beq
16\pi^2 \frac{dm_S^2}{d\log\mu^2} = 3\left|\lambda_Q\right|^2 \left(m_S^2 + m_{Q_1}^2 + m_{Q_2}^2 + \left|A_{\lambda_Q}\right|^2\right),
\label{eq:SsoftRG}
\eeq
shows that for $\left<S\right> < M_{mess}$, RG flow drives $S$ tachyonic and leads to a nonzero VEV that will be determined by an interplay between the SUSY-breaking potential (which wants $S$ to be large) and the tree-level $|F_Y|^2$ potential (which wants $S$ to be small). There are two different scenarios for the parametric behavior of the potential, depending on the relationship between PQ and messenger scales:
\begin{itemize}
\item The PQ scale is smaller, $\left<S\right> \simlt M_{mess}$. We assume $M_{mess}$ is not {\em much} larger than the PQ-breaking scale. In this case we can use Eqs. \ref{eq:MGMsquarkmass} and \ref{eq:SsoftRG} to estimate
\beq
m_S^2(S) \approx -\frac{\alpha_s^2 \lambda_Q^2}{8\pi^4} \frac{F^2}{M_{mess}^2} \log\frac{M_{mess}}{\lambda_Q S},
\eeq
and minimizing the potential $V(S)\approx m_S^2 S^\dagger S + |F_Y|^2$ gives:
\beq
\left<\left|S\right|\right>^{2n+2} = \frac{\alpha_s^2}{8(n+2)\pi^4} \left|\frac{\lambda_Q}{\lambda_Y}\right|^2 \left(\frac{F M_P^n}{M_{mess}}\right)^2 \log \frac{M_{mess}}{\lambda_Q \left<S\right>}.
\eeq
Crudely, then, the relationship between the PQ-breaking scale and the SUSY-breaking scale in this model is:
\beq
f_a \sim \left( \frac{FM_P^n}{M_{mess}} \right)^{\frac{1}{n+1}}\sim \left(m_{soft} M_P^n\right)^{\frac{1}{n+1}}.
\eeq
For $n=1$, the Kim--Nilles $\mu$ term $\frac{f_a^2}{M_P}$ is automatically of the same order as the soft masses. As a numerical example, if we fix $n=1$, $\lambda_Q = \lambda_Y = 1$ and $m_{Q_i}^2 = \left(1~{\rm TeV}\right)^2$, we find that $M_{mess} = 10^{12}$ GeV corresponds to $\left<S\right> = 5.2\times10^{10}~{\rm GeV}$. This particular choice corresponds to $\sqrt{F} = 2.8 \times 10^8~{\rm GeV}$ and $m_{3/2} = 4.4~{\rm MeV}$. In general, this scenario favors $m_{3/2} \sim {\cal O}(100~{\rm keV})$ or somewhat larger, corresponding to relatively high-scale gauge mediation and stable NLSPs on collider time scales. Because the scenario is discussed more extensively in reference \cite{Choi:2011rs}, we will focus more on the following scenario, which allows lighter gravitino masses.
\item The PQ scale is larger, $\left<S\right> \gg M_{mess}$ (for calculability, we assume a hierarchy). In this case the correct calculation of the potential appears in Section 5.2 of \cite{ArkaniHamed:1998kj} as an example of analytic continuation into superspace. Assuming that QCD effects are dominant, and keeping only the largest power of $\log\frac{S}{M_{mess}}$, their result is:
\beq
\left|S\right| \frac{\partial V_{\rm eff}}{\partial \left|S\right|} = -\frac{\alpha_s^2}{12 \pi^4} \left|F\right|^2 \log^2\frac{\lambda_Q \left|S\right|}{M_{mess}}.
\label{eq:dVdS}
\eeq
Minimizing the combination of this and the $|F_Y|^2$ tree-level potential, we find:
\beq
\left<\left|S\right|\right>^{2n+4} = \frac{\alpha_s^2}{24(n+2)\pi^4}\left|\frac{F M_P^n}{\lambda_Y}\log \frac{\lambda_Q \left|S\right|}{M_{mess}}\right|^2.
\label{eq:PQvSUSY}
\eeq
Thus, schematically, in this example we have a scaling:
\beq
f_a \sim \left(F M_P^n\right)^{\frac{1}{n+2}},
\eeq
so that even though the establishment of the PQ-breaking VEV depends on SUSY-breaking dynamics, the scale of PQ breaking is much larger than the scale of SUSY breaking. This allows low-scale SUSY breaking, with NLSP decays to gravitino on collider timescales, to be consistent with axions in the allowed window for conventional axion cosmology. 
\end{itemize}

\FIGURE[!h]{
\includegraphics[width=0.475\textwidth]{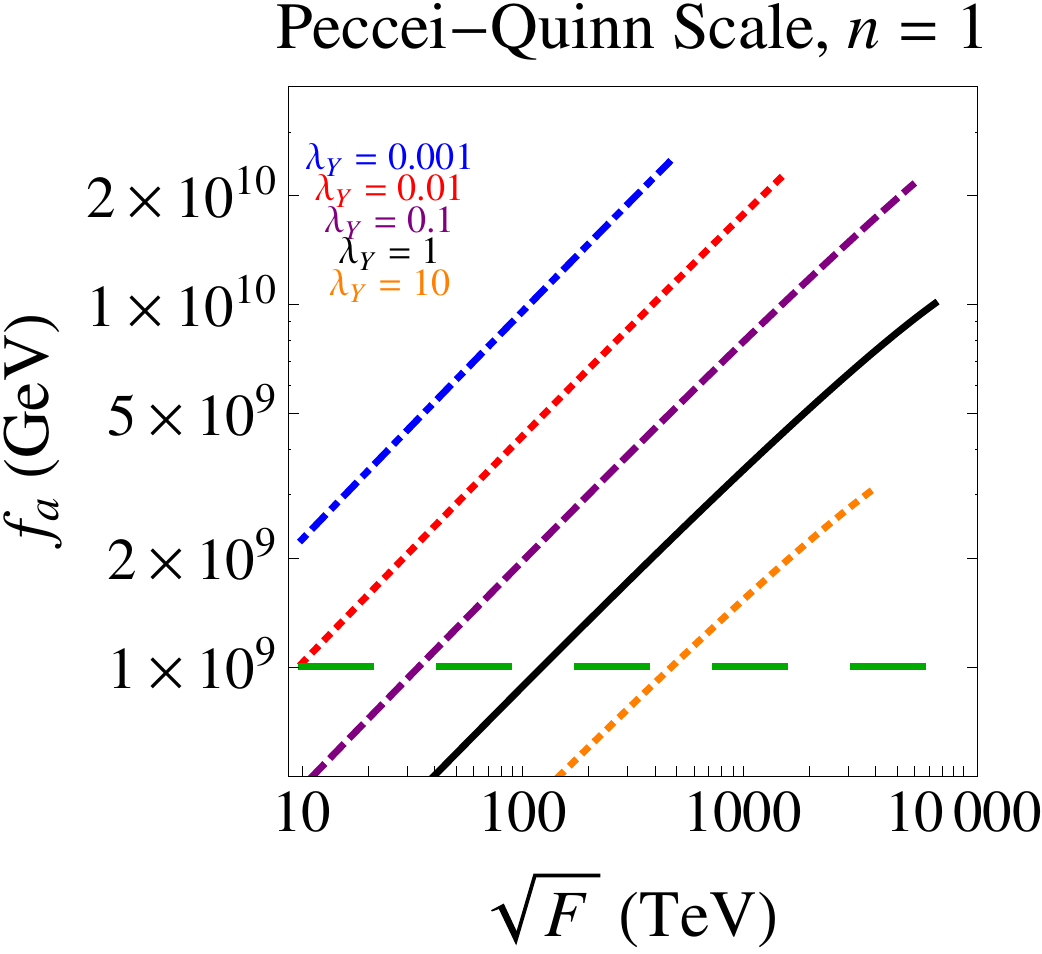}~~\includegraphics[width=0.48\textwidth]{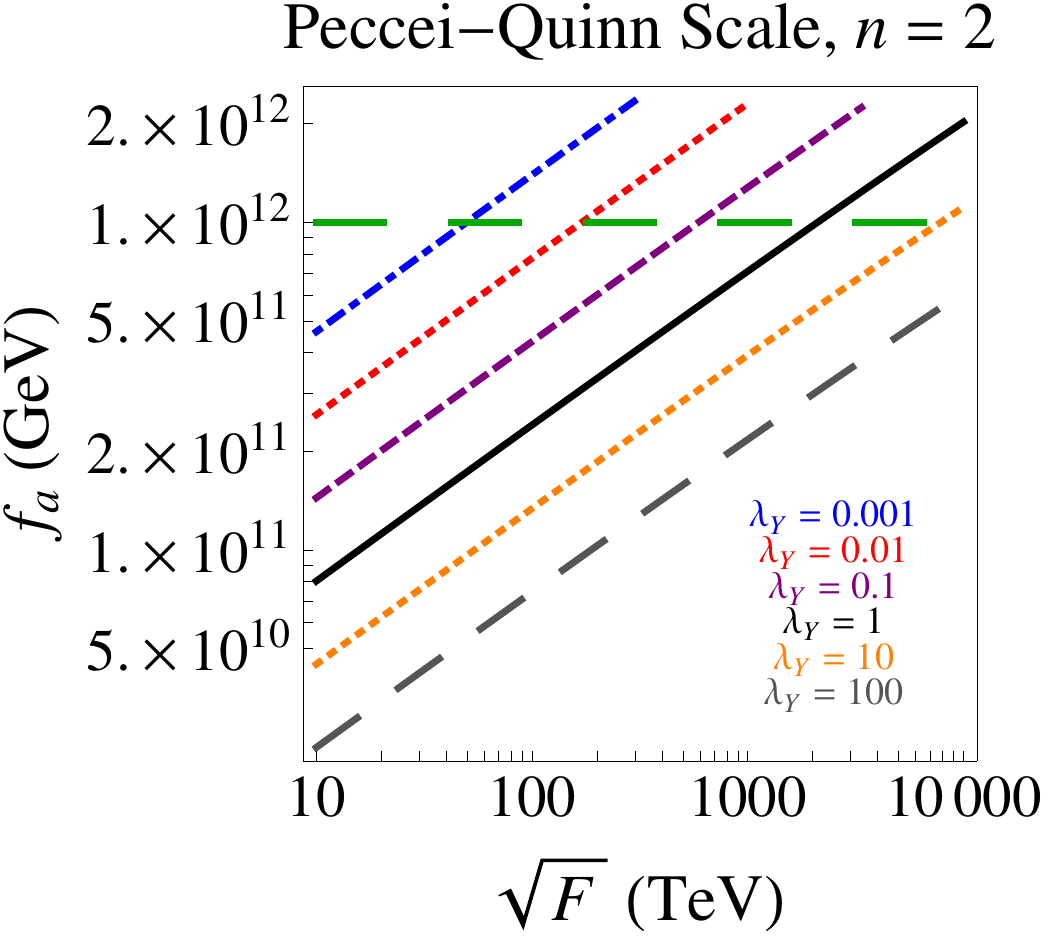}
\caption{The axion decay constant $f_a$ versus the SUSY-breaking scale $\sqrt{F}$, based on Eq. \ref{eq:PQvSUSY}. To fix $M_{mess}$, we require that the squark soft masses (Eq. \ref{eq:MGMsquarkmass}) are 1 TeV. The coupling $\lambda_Q = 1$ for all curves (because it appears only in the log, varying it has little effect.) The (from bottom to top) long dashed grey line, dotted orange line, solid black line, dashed purple line, dotted red line, and dot-dashed blue line are for $\lambda_Y$ equal to $100$, $10$, $1$, $0.1$, $0.01$,  and $0.001$, respectively. The horizontal green dashes at 10$^9$ GeV signal the lower bound on $f_a$ consistent with stellar cooling, while those at 10$^{12}$ GeV mark the bound at which axions overclose the universe.}
\label{fig:PQvsSUSY}}

We will now turn to a more detailed study of the latter case. In Figure \ref{fig:PQvsSUSY}, we illustrate the relationship between the SUSY-breaking scale, $\sqrt{F}$, and the axion decay constant $f_a$. For $n=1$, it is not possible to extrapolate to larger $\sqrt{F}$ and $f_a$. The reason is that, if we want soft masses to be in the TeV regime, $M_{mess}$ must be taken proportional to $F/m_{soft}$, and for $\sqrt{F} \simgt 10^4~{\rm TeV}$, $M_{mess}$ becomes larger than $\left<S\right>$ and the model enters the higher scale SUSY-breaking regime studied by  \cite{Choi:2011rs}. From the left panel in Figure \ref{fig:PQvsSUSY}, the accessible range of $f_a$ for $n=1$ is roughly $10^9~{\rm GeV} \simlt f_a \simlt 10^{10}~{\rm GeV}$, and achieving such values of $f_a$ for low-scale SUSY breaking is easier at small $\lambda_Y$. (Of course, in general, one could interpolate between the two regimes; there is no {\em physical} bound on $\sqrt{F}$ or $f_a$, just a region in which neither approximation we have described is quite accurate.) For $n=2$, the scaling $f_a \sim (F M_P^2)^{1/4}$ allows for a larger separation between the PQ scale and the SUSY breaking scale. As shown in the right panel in Figure \ref{fig:PQvsSUSY}, our model with $n=2$ probes the region $10^{10}~{\rm GeV} \simlt f_a \simlt 10^{12}~{\rm GeV}$, at the upper end of which the axion could be cold dark matter. In this case, large $\lambda_Y \simgt \calo(1)$ are preferred as for fixed $\sqrt{F}$, small $\lambda_Y$ could push $f_a$ above the upper limit.  

\subsection{Saxion Mass and Decays}

\FIGURE[!h]{
\includegraphics[width=0.46\textwidth]{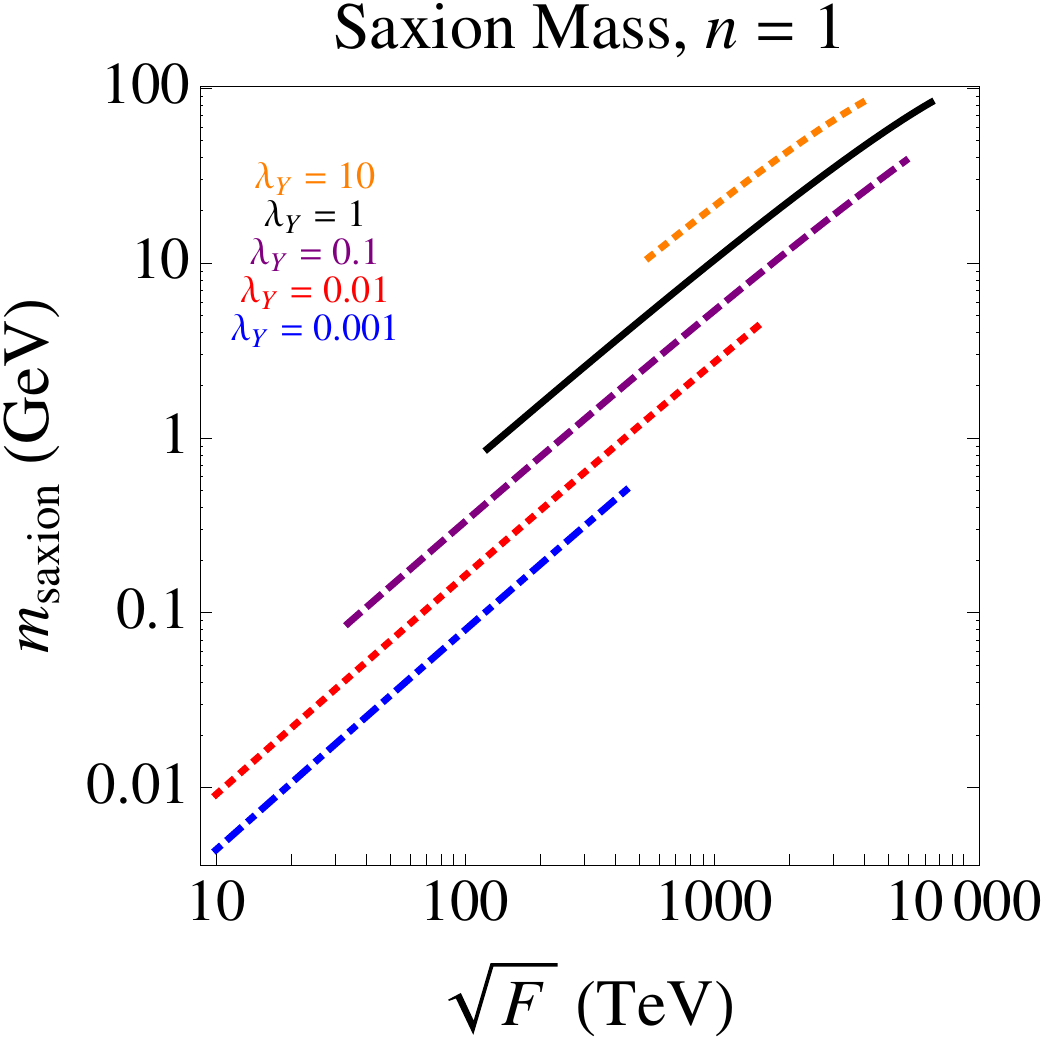}~\includegraphics[width=0.47\textwidth]{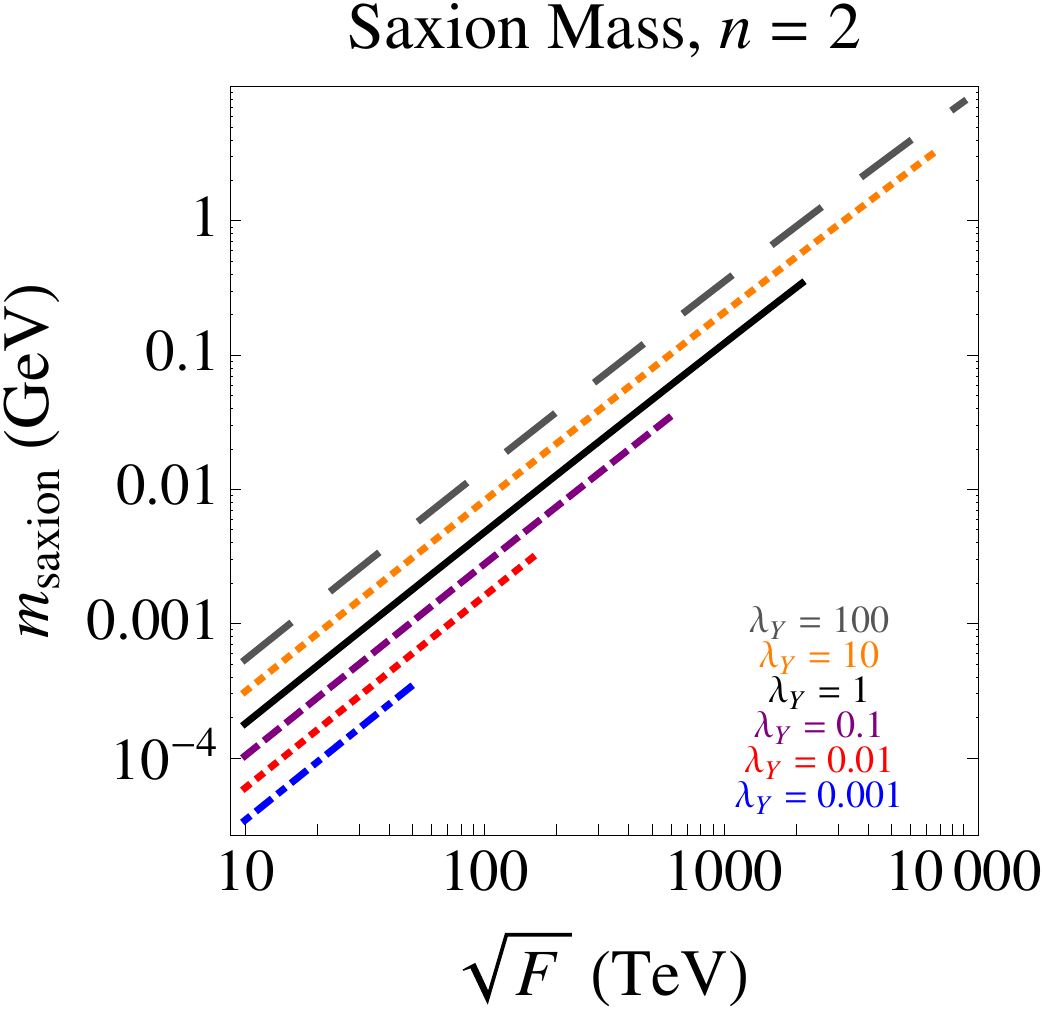}
\caption{Saxion mass versus the SUSY-breaking scale $\sqrt{F}$, based on Eq. \ref{eq:msaxion}. Parameters are as in Figure \ref{fig:PQvsSUSY}, with (from top to bottom) long dashed grey line, dotted orange line, solid black line, dashed purple line, dotted red line, and dot-dashed blue line are for $\lambda_Y$ equal to $100$, $10$, $1$, $0.1$, $0.01$,  and $0.001$. Points are plotted {\em only} if 10$^9$ GeV $< f_a < 10^{12}$ GeV for the given parameter choices.}
\label{fig:SaxionMass}}

In the low-scale SUSY breaking regime discussed above, it is straightforward to calculate that the saxion mass around its minimum is:
\beq
m_s^2 = \frac{\alpha_s^2 F^2}{12 \pi^4 \left<S\right>^2} \left((n+2)\log^2\frac{\lambda_Q \left<S\right>}{M_{mess}} - \log\frac{\lambda_Q \left<S\right>}{M_{mess}}\right).
\label{eq:msaxion}
\eeq
As illustrated in Figure \ref{fig:SaxionMass}, the resulting masses can range from MeV to 100 GeV for the model choices that give us low-scale SUSY breaking with reasonable axion decay constants. Because couplings are $1/f_a$ suppressed, there is no bound on such particles even when they are light.

Saxions always have a decay to a pair of axions, which arises from the kinetic term:
\beq
\Gamma(s \to aa) = \frac{m_s^3}{32\pi f_a^2}.
\eeq
This will lead to a disaster for cosmology if it is the dominant decay mode: because the saxion dominates the energy density as it oscillates coherently around its minimum, if it decays predominantly to axions the universe will thereafter be full of highly relativistic particles that do not thermalize with SM degrees of freedom. We must require that the dominant decay of the saxion is to SM fields. Note that the decay to two gluons from the operator $SW_\alpha W^\alpha$ does not help, as the associated partial width is smaller by a factor of $\left(\frac{\alpha_s}{4\pi}\right)^2$ relative to the partial width to two axions.

This is where the $S^2 H_u H_d$ superpotential term plays its key role, providing the crucial mechanism for saxion decays to lead to reheating. The mass mixing term for Higgs and saxion (taking the decoupling limit for the MSSM Higgses) is:
\beq
{\cal L} \supset \frac{4 \mu^2 v}{\left<S\right>} s h.
\eeq
(In particular, $\tan \beta$ dependence drops out.) For instance, if $m_s > 2 m_b$, we have a decay $s \to b\bar{b}$ with partial width:
\beq
\Gamma(s \to b\bar{b}) = \frac{3 \mu^4}{\pi \left<S\right>^2 m_h^4}m_s m_b^2 \left(\sqrt{1 - \frac{4 m_b^2}{m_s^2}}\right)^3 \sim 96 \left(\frac{\mu}{m_h}\right)^4 \left(\frac{m_b}{m_s}\right)^2 \Gamma(s \to aa).
\eeq
For smaller $m_s$, e.g., $2m_c<m_s<2m_b$, the saxion could decay to charmed mesons (or $\tau$'s). For $\mu \simgt 2 m_h$, the saxion would decay to the SM $99\%$ of the time and thus the energy density contained in relativistic axions does not spoil BBN. Near the upper bound of $f_a$, $10^{11}$ GeV $\simlt f_a \simlt 10^{12}$ GeV, the region that our model with $n=2$ covers, the reheating temperature, $T_{RH}$, is lower, ranging from a few MeV to 100 MeV. On the other hand, near the lower bound of $f_a$, $10^9$ GeV $\simlt f_a \simlt 10^{10}$ GeV, the region that our model with $n=1$ covers, $T_{RH}$ could be as large as  $\calo(10)$ GeV.

\subsection{Cosmology}

We wish to dilute the abundance of moduli fields, which we assume have a mass $m_T$ of approximately the order of $m_{3/2}$ (but perhaps somewhat larger). Specifically, we will write:
\beq
m_T \equiv \xi \frac{F}{M_P},
\eeq
where for instance in KKLT $\xi \sim \log\frac{M_P}{m_{3/2}}$ can be $\sim 10^2$. The universe rapidly becomes matter-dominated for $3H \simlt m_T$. As explained in Ref. \cite{Randall:1994fr}, at this point we have:
\beq
\frac{n_\hatt}{s_{\rm rad}} = \frac{\frac{1}{2}m_T t_0^2}{\frac{2\pi^2}{45}g_{*s}T_{\rm M.D.}^3} \sim 0.05 \frac{t_0}{\sqrt{\xi F}},
\label{eq:modulitoentropy}
\eeq
where we take the effective number of degrees of freedom to be $g_{*s} =275$ and the moduli oscillation amplitude $t_0 \sim M_P$. $n_\hatt/s_{\rm rad}$ remains fixed in the absence of entropy production. The temperature of the radiation at the point when the moduli dominate the energy density is $T_{\rm M.D.} \approx (90/(\pi^2g_{*s}))^{0.25} \sqrt{\xi F} \approx 0.4 \sqrt{\xi F}$. Due to the $t_0/\sqrt{F}$ factor, for low-scale SUSY breaking $\sqrt{F} \sim (10 - 10^4)$ TeV, $n_\hatt/s$ can be as large as $(10^{9} - 10^{12})$. Of course, the number that must be small at late times is $\rho_\hatt/s$, which shrinks with $m_T$, and can be estimated to be $\sim (10^3 - 10^6)$ GeV. Given the critical density of the Universe versus the present entropy is $\rho_c/s = 3.6 \times 10^{-9} h^2$ GeV, where $h$ is the Hubble rate, the moduli density has to be diluted by a factor of $\sim (10^{12} - 10^{15})$!

Let's consider how the saxion can help to solve this problem by realizing the idea of thermal inflation \cite{LythStewart}. By integrating Eq. \ref{eq:dVdS}, and assuming the cosmological constant is approximately zero at the PQ-breaking zero-temperature vacuum, we see that the vacuum energy when $S = 0$ is approximately given by:
\beq
V_0 \equiv V(S = 0) \approx \frac{\alpha_s^2}{36 \pi^4} \left|F\right|^2 \log^3\frac{\lambda_Q f_a}{M_{mess}} \sim \left((0.1- 0.4)~\sqrt{F}\right)^4,
\eeq
where the range of numbers quoted is extracted from scanning the set of model lines displayed in Fig. \ref{fig:PQvsSUSY}. Thus, we expect the vacuum energy for $S$ at the origin corresponds to a scale slightly, but not far, below $\sqrt{F}$. Thus, when the Hubble scale drops to order $\sqrt{F}$, this constant energy density provides a potential source of late inflation. Because we have $V_0^{1/4} \sim 0.1 \sqrt{F}$, and $T_{\rm M.D.} \sim \sqrt{F}$, moduli domination and the onset of thermal inflation happen almost at the same time, but by assuming that $\xi$ is somewhat large we will simplify the discussion and speak as if events are well-separated in time. Thermal inflation begins when $V_0$ dominates the energy density, i.e. $V_0 \geq m_T n_\hatt$, which occurs when the radiation is at a temperature:
\beq
T_{\rm T.I.} \approx \left( \frac{30V_0}{\pi^2 g_{*s}T_{\rm{M.D}}}\right)^{1/3} \sim (0.02-0.06)~\sqrt{F}.
\eeq

In order for $V_0$ to lead to inflation, we need $S$ to remain at the origin for a period of time. But this happens automatically in our model: at high temperatures, the saxion field will be trapped at the origin by thermal corrections to its potential. These corrections arise from interactions with the $Q$ fields, which keep it in thermal equilibrium with SM degrees of freedom. In particular, the leading thermal term in the potential goes as \cite{Dolan:1973qd}:
\beq
V_{\rm thermal}(S) = \frac{3}{4} \left|\lambda_Q\right|^2 T^2 \left|S\right|^2.
\label{eq:VthermalS}
\eeq
Thus $S$ remains at the origin until the universe cools enough that the tachyonic mass from eq. \ref{eq:SsoftRG} becomes larger. Then, $S$ begins to roll out to its VEV when the temperature drops to a scale set by the soft masses of the $Q$s:
\beq
T_{\rm roll} \approx \frac{m_Q}{\pi} \sqrt{\log\frac{M_{mess}}{m_Q}} \sim 0.5~{\rm to}~1~{\rm TeV}.
\eeq

During the thermal inflation, the number density of $\hatt$ redshifts as $e^{-3Ht}$ while the temperature $T$ redshifts as $e^{-Ht}$. Thus we have (using temperature as a clock):
\beq
n_\hatt(T_{\rm roll}) = n_\hatt(T_{\rm T.I.}) \left(\frac{T_{\rm roll}}{T_{\rm T.I.}}\right)^3.
\eeq
In particular, the ratio of $\rho_\hatt = m_T n_\hatt$ to the total energy density changes by this factor, because the total energy density remains approximately $V_0$ throughout thermal inflation.

After thermal inflation, the energy $V_0$ is converted to energy $m_S n_S$ for the saxion $s$ oscillating around its minimum. The ratio $n_\hatt/n_S$ remains fixed as $s$ oscillates:
\beq
\frac{n_\hatt}{n_S} = \frac{m_S}{m_T} \frac{\rho_\hatt}{\rho_S} \sim \frac{m_S}{m_T} \left(\frac{T_{\rm roll}}{T_{\rm T.I.}}\right)^3.
\eeq
When $s$ decays, the entropy changes to
\beq
s_{\rm rad} = \frac{2\pi^2}{45} g_{*s} T_{RH}^3 = \frac{4}{3} \frac{\rho_{\rm rad}}{T_{RH}} = \frac{4}{3} \frac{m_S n_S}{T_{RH}},
\eeq
with $T_{RH}$ the reheating temperature of $s$. From this one can conclude that the final value of the energy density in the modulus versus the entropy, in terms of the temperatures, is:
\beqs
\frac{\rho_\hatt}{s_{\rm rad}} &=& m_T \frac{n_{\hatt}}{n_S} \frac{n_S}{s_{\rm rad}} \sim m_T \frac{m_S}{m_T} \left(\frac{T_{\rm roll}}{T_{\rm T.I.}}\right)^3  \frac{3 T_{RH}}{4 m_S} \approx \frac{3}{4} T_{RH} \left(\frac{T_{\rm roll}}{T_{\rm T.I.}}\right)^3, \nonumber \\
\Omega_\hatt h^2& =& \frac{\rho_\hatt/s_{\rm rad}}{\rho_c/s_{\rm rad}} \approx 2 \times 10^8~\frac{\rho_\hatt}{s_{\rm rad}}\approx 2 \times 10^8 ~\frac{T_{RH}}{{\rm GeV}} \left(\frac{T_{\rm roll}}{T_{\rm T.I.}}\right)^3
\eeqs
Thus, we want to lower $T_{RH}$ while making the ratio $T_{\rm roll}/T_{\rm T.I.}$ as small as possible. It turns out that for relatively large SUSY breaking scale $\sqrt{F} \simgt 2\times10^3$ TeV and low reheating temperature $T_{RH} \sim 10$ MeV, the saxion oscillation and decay could reduce the moduli density to be below the critical density. This implies that only our model with $n=2$ works, where $f_a$ could be close to $10^{12}$ GeV, hence the saxion mass is relatively small, and consequently $T_{RH}$ could be as low as 10 MeV.

As briefly mentioned in the introduction, for a modulus with mass in the range 200 keV to 10 GeV, even if its lifetime is longer than the age of the Universe, its decays to photons in the past would contribute to the $\gamma$-ray flux~\cite{Kawasaki:1997ah,Hashiba:1997rp,Asaka:1997rv}. Thus the observed $\gamma$-ray background provides a much more stringent constraint on the moduli density, as shown in Fig. 1 in~\cite{Hashiba:1997rp}. The bound gets strongest for moduli with mass around 100 MeV, requiring $\Omega_\hatt h^2<10^{-10}$! As this constraint is obtained by assuming that moduli dominantly decay to two photons through the operator $c T F^{\mu\nu}F_{\mu\nu}$ with $c$ a model-dependent coupling, it might be relaxed if $c$ is suppressed, e.g., by the gauge coupling $c \sim \alpha_{em}/(4\pi)$, because the moduli lifetime becomes longer and fewer have decayed at a given time. Considering that the ultraviolet theory could possess many moduli, one would still need to worry that at least one modulus will have unsuppressed width to photons. With this in mind, assuming this constraint applies (at least to some of the moduli), one would find several interesting implications. First, in the KKLT moduli stabilization, the modulus mass is fixed at roughly 100 times the gravitino mass; for $\sqrt{F} \simgt 2\times 10^3$ TeV, $m_T \simgt$ 200 keV, and thermal inflation, though capable of reducing the moduli density to just below the critical density, could not dilute it sufficiently to satisfy the constraint from the $\gamma$-ray flux. One has to invoke a more efficient dilution mechanism, or adopt a model of moduli stabilization with a smaller ratio of modulus to gravitino mass. This constraint also applies to relatively high scale mediation with $\sqrt{F}>10^4$ TeV as considered in~\cite{Choi:2011rs}. The region of parameter space under the least tension from the various constraints has $\sqrt{F} \approx  2\times10^3~{\rm to}~10^4~{\rm TeV}$, $T_{RH} \approx 10$ MeV, and moduli masses $m_T \approx m_{3/2}$ without the log-enhancement as in KKLT.

\section{Discussion}
\label{sec:discuss}
\setcounter{equation}{0}
\setcounter{footnote}{0}

We have argued that, generically, low-scale SUSY breaking suggests the existence of light moduli fields with masses approximately of order $m_{3/2}$. One of the most interesting implications is that the universe likely had a period of late entropy production. If we evade the generic scenario by tuning the moduli to be heavy enough that their decays reheat the universe to above 10 MeV and allow for BBN, the entropy production arises from the decay of the moduli fields themselves. If we aim for a more natural scenario, the entropy production must come from some other source; we have suggested the saxion as a natural candidate. In the case of gravity mediation with $m_{3/2}$ of weak scale, which we have not considered here, one can similarly argue that a reasonable scenario is that moduli are at around 100 TeV and their decays reheat the universe (for several incarnations of this idea, see Ref. \cite{NonthermalGravityMediation}). 

The big picture is that considerations of moduli cosmology {\em generically} lead us to expect an association between SUSY and low reheating temperatures, for any scale of SUSY-breaking that could lead to weak-scale superpartners. Another line of reasoning, completely independent of moduli problems, has recently led to a similar conclusion. Namely, the combination of gravitino cosmology \cite{Moroi:1993mb} and the cosmology of a QCD axino, considered in tandem, require low reheating temperatures~\cite{Cheung:2011mg}.\footnote{This follows earlier work on axino cosmology, including the possibility of axino dark matter and bounds on reheating temperature~\cite{Rajagopal:1990yx, CoviAxinos, Strumia:2010aa}. See Ref.~\cite{Bae:2011jb}, however, for an argument that in particular models, especially of the DFSZ type, this bound might be circumvented. Further recent work on cosmology of gravitinos, axinos, and saxions appears in Ref.~\cite{HasenkampKersten}.}

One exciting possibility suggested by these considerations is that there is a great deal of physics happening at scales that are not far away. This potentially extends beyond the saxion physics and the SUSY-breaking sector. For instance, given low reheating temperatures, one might expect that baryogenesis happens at low scales, even below the electroweak phase transition, and perhaps can be probed at colliders. Scenarios such as those of Ref. \cite{LowScaleBaryogenesis} potentially fit into the larger picture of cosmology that we have sketched. An alternative possibility, advocated in \cite{de Gouvea:1997tn}, is Affleck-Dine baryogenesis that produces such a large baryon asymmetry that it can explain the observed number even {\em after} being diluted by late entropy production. However, such a scenario deserves a careful check in light of possible $Q$-ball dangers \cite{Berkooz:2005sf}. 

Though we have focused mostly on the cosmology of low-scale SUSY breaking, we want to mention a possible collider signature related to the saxion scenario. There in some cases (e.g., $n=1$ models), the superpartner of the saxion, the axino, has mass at or below the weak scale, $m_{\tilde{a}} \simlt \mu$, and mixes with the Higgsinos through the superpotential operator $S^2H_uH_d$. Thus it's possible to have the NLSP (e.g., Higgsinos) decaying to axino plus $Z$ or Higgs with collider length scale~\cite{Martin:2000eq,Nakamura:2008ey}. The signature can mimic signatures previously studied in the GMSB context \cite{MatchevThomas,NLSPPromptPaper,NLSPLongLivedPaper}. However, for the axino decay mode to be comparable to that to gravitino, the PQ scale has to be relatively low, $f_a \simlt 10^{10}$ GeV, which doesn't work well for the saxion being a late-time entropy production source. Yet this collider signature could well be present for relatively high scale SUSY breaking and mediation. It raises the interesting question: suppose we find evidence for a neutralino that propagates over a macroscopic distance and decays to a $Z$ or a Higgs boson plus an invisible light particle. Should we interpret it as a decay to a gravitino, an axino, or something else? Such decays offer the prospect of measuring important scales like $\sqrt{F}$ or $f_a$, but only if the correct interpretation is found.

In summary, we see many interesting possibilities for consilience between TeV-scale particle physics, including accessible collider signals, and the fundamental physics of the Planck scale, driven by the cosmological problems posed by very weakly interacting particles. It would be interesting to pursue more complete models, including cases with multiple moduli, and to compute the cosmology of saxion-driven thermal inflation in more detail.

\section*{Acknowledgments}

We thank Ian-Woo Kim, Masahito Yamazaki, and especially Josh Ruderman for involvement in an early stage of this work. We thank David Shih for useful comments on the manuscript. We would also like to thank Nima Arkani-Hamed, Zackaria Chacko, Nathan Seiberg, and Tomer Volansky for useful comments, suggestions, and criticisms. J.F. is supported by the DOE grant DE-FG02-91ER40671. M.R. thanks the PCTS for its support. L.T.W. is supported by the NSF under grant PHY-0756966 and the DOE Early Career Award under grant DE-SC0003930. 

\appendix
\section{Review: Origin and Properties of Moduli Fields}
\label{app:modprop}

In string or M theory, one finds that parameters like gauge couplings are never fixed numbers, but are determined by the expectation value of scalar fields. For instance, if a gauge field in the 4D effective theory at low energies arises from D7 branes wrapping a 4-dimensional cycle $\Sigma$ in a Calabi-Yau manifold, the low-energy gauge coupling is determined by the volume of that cycle, which is a dynamical quantity because the theory includes gravity. More precisely, $\frac{1}{g^2} \propto \rm{Vol}(\Sigma)$. In a supersymmetric theory, this is the real part of a chiral superfield with an imaginary part realted to the $\theta$ angle of the theory, $\theta \propto \int_\Sigma C_4$, where $C_4$ is the Ramond-Ramond 4-form of type IIB string theory. This is a representative example of a wide class of moduli in string theory which have approximate continuous axionic shift symmetries, $\theta \to \theta + \delta \theta$, broken to discrete symmetries (e.g., $\theta \to \theta + 2\pi$) by instantons. Examples are K\"ahler moduli in IIB strings and both K\"ahler and complex structure moduli in IIA strings. Moduli that determine the values of gauge couplings are always of this form, because there is always an associated $\theta$-angle. This means, in particular, that any superpotential terms involving such moduli are of the form $Ae^{-b T}$, with possible values of the number $b$ constrained by the remaining discrete shift symmetry.

Because the moduli arise from closed string modes like the graviton, Ramond-Ramond fields, or $B$ field, they couple to normal matter with gravitational strength (i.e., through $1/M_{Pl}$-suppressed operators). While we know that our universe doesn't contain massless scalars with gravitational-strength interactions, fifth-force searches still allow fairly light fields with such couplings to exist. At the very lowest possible scales of SUSY breaking, $\sqrt{F} \sim 10$ TeV, direct searches for these forces may be relevant~\cite{Dimopoulos:1996kp}, with current experimental bounds excluding gravitational-strength forces at a range $\lambda = 56~\mu{\rm m} = \left(3.5~{\rm meV}\right)^{-1}$~\cite{Kapner:2006si}.

One other property of moduli that will be of interest for us is that they generally have noncanonical K\"ahler potentials that diverge as moduli approach certain special points. For instance, in the D7-brane gauge coupling example, the limit where $\rm{Vol}(\Sigma) \to 0$ is associated with the existence of a proliferation of new light states (from objects wrapping the cycle that become massless when it collapses). In field theory, the procedure of integrating out these states is no longer valid when they become massless, so we should expect that approaching this point in the moduli space leads to divergences. In the opposite limit, ${\rm Vol}(\Sigma) \to \infty$, KK modes become light, and a similar divergence occurs. In our examples, this will mean that moduli have logarithmic K\"ahler potentials.

Further explanation of the origin of moduli fields in string compactifications may be found in the textbook \cite{PolchinskiBook}. An interesting discussion of generic properties of string theory moduli spaces that are conjectured to be a requirement of all consistent theories of quantum gravity is found in Ref. \cite{Ooguri:2006in}.

\end{document}